\documentclass[font=lmodern, cite=numeric]{labpaper}

\usepackage[utf8]{inputenc}
\usepackage{mathtools}
\usepackage{algorithm}
\usepackage[noend]{algpseudocode}
\usepackage{array}
\usepackage{xurl}
\usepackage[capitalise,noabbrev]{cleveref}
\usepackage{enumitem}
\usepackage{float}
\hypersetup{
  pdftitle={Structure Matters: Differentially Private Spectral Statistics for ECG Time Series},
  pdfauthor={Reza Saadi, Erfan Salavati, Mehmet Cengiz Onbaşlı}
}
\newcommand{\doi}[1]{doi: \href{https://doi.org/#1}{{\urlstyle{rm}\def\UrlBreaks{\do\/}\def\UrlBigBreaks{}\nolinkurl{#1}}}}

\newif\ifshowrev \showrevfalse
\newcommand{\rev}[1]{\ifshowrev\textcolor{red}{#1}\else#1\fi}

\newcommand{\R}{\mathbb{R}}
\newcommand{\E}{\mathbb{E}}
\newcommand{\cN}{\mathcal{N}}
\newcommand{\cM}{\mathcal{M}}

\newcommand{\eps}{\varepsilon}
\newcommand{\deltap}{\delta}
\newcommand{\norm}[1]{\left\lVert #1 \right\rVert_2}
\newcommand{\clip}{\operatorname{clip}}
\newcommand{\PSD}{\operatorname{PSD}}
\newcommand{\argmin}{\operatorname*{arg\,min}}

\newcommand{\fullDim}{d}
\newcommand{\projDim}{k}
\newcommand{\groupSize}{m}
\newcommand{\clipRad}{C}
\newcommand{\noiseMult}{\sigma_{\eps,\deltap}}
\newcommand{\noiseStd}{\tau}

\newcommand{\dctMat}{U}
\newcommand{\selector}{P_{\projDim}}
\newcommand{\windowVec}{x}
\newcommand{\statVec}{s}
\newcommand{\projVec}{z}
\newcommand{\clippedVec}{\bar z}
\newcommand{\releasedVec}{\widetilde z}
\newcommand{\releasedStat}{\widetilde s}
\newcommand{\targetMean}{\mu}
\newcommand{\query}{q_{k,C}}

\newcommand{\codefreeze}{18 August 2026}

\firstlogo{figures/logo-university}

\logoheight{1.5cm}


\title{Structure Matters: Differentially Private Spectral Statistics for ECG Time Series}

\addaffiliation{1}{Koç University, Istanbul, Türkiye}
\addaffiliation{2}{Independent Researcher}

\addauthor[affil=1, email=hdadmarzi23@ku.edu.tr, corresponding]{Reza Saadi}
\addauthor[affil=2, email=erfan.salavati@gmail.com]{Erfan Salavati}
\addauthor[affil=1, email=monbasli@ku.edu.tr]{Mehmet Cengiz Onbaşlı}

\begin{document}
\maketitle
\makeatletter
\begingroup
  \let\thefootnote\relax
  \long\def\@makefntext#1{\noindent#1}%
  \footnotetext{An earlier version of this paper was accepted at the NeurIPS 2026 Workshop on Generalization for Time Series in Tight Settings (TS-LIMITS). This version adds a post-hoc ablation (\cref{sec:ablation,app:ablation}) \rev{and clarifying text}.}%
\endgroup
\makeatother

\begin{abstract}

Releasing high-dimensional spectral summaries under differential privacy (DP) introduces large error when every coordinate is perturbed. We ask whether frequency structure helps beyond dimension reduction. We release the mean Welch log-power spectral density (log-PSD) of groups of electrocardiogram (ECG) windows under window-level DP by keeping a publicly calibrated low-order discrete cosine transform (DCT) prefix, then clipping, averaging and adding Gaussian noise. With every setting fixed before any test record was downloaded, on 16 untouched MIT-BIH records (groups of 40 windows, $(\eps,\deltap)=(1,10^{-5})$) the structured release cuts median relative $L_2$ error by $39.6\%$ and band error by $51.8\%$ against equal-size random DCT masks with independently calibrated radii, and relative $L_2$ error by $78.5\%$ against full-dimensional release, close to the $78\%$ our error decomposition predicts. A post-hoc ablation traces most of this gain to the constant DCT mode that carries the curve's overall level. For private spectral summaries, which coordinates are released matter and not only their counts. Choosing the released subspace on public data to match the structure of the statistic can therefore cut error substantially at no extra privacy cost.

\end{abstract}

\section{Introduction}
\label{sec:introduction}

Power spectral density (PSD) curves describe periodic structure and the distribution of signal energy across frequencies. They provide useful summaries for time-series analysis without requiring release of the raw waveform, but they do not by themselves protect the underlying measurements. Differential privacy (DP) provides a formal guarantee for such releases that is preserved under arbitrary post-processing. Its utility cost, however, can be large when noise is added to every coordinate of a long spectral vector~\citep{dworkroth2014}.

Transform-based DP mechanisms address this cost by concentrating a signal in a smaller set of coefficients. The Fourier Perturbation Algorithm is an early example for aggregate time-series queries~\citep{rastogi2010fpa}. Other work protects spectral properties of stochastic processes~\citep{parker2022spdp}, sanitizes individual electrocardiogram (ECG) signals under metric privacy using a discrete cosine transform (DCT) representation~\citep{bonomi2022ecg}, or perturbs neural-network gradients in a spectral domain~\citep{feng2023spectraldp}. These settings differ in what is private and what is released. We focus on a specific question for aggregate spectral statistics: \emph{does preserving frequency structure improve utility beyond retaining fewer coordinates?}

We use ECG recordings as a real-world benchmark for this question. Each window is mapped to a Welch estimate of the logarithm of its PSD (log-PSD curve), whose coordinates are ordered by physical frequency. We apply an orthonormal DCT, retain a low-order prefix, clip each projected vector, average over a group of windows, and add Gaussian noise. These low-order modes describe smooth variation across the log-PSD frequency axis; they are not a selection of the lowest physical frequencies in the ECG signal. Random subsets of the same number of DCT coefficients provide controls for dimension reduction without this preference for smooth structure.

The paper makes three contributions.
\begin{itemize}[leftmargin=1.4em]
  \item We give a DCT-prefix mechanism for aggregate log-PSD curves with a window-level DP guarantee under replace-one adjacency. We also provide an exact error decomposition that accounts for projection, clipping, and Gaussian noise.

  \item We compare the structured mechanism with full-dimensional DP, a dimension- and noise-matched random-DCT control, and five fixed random DCT masks with independently calibrated clipping radii. The projection dimension and clipping radii are selected using public records only. Development follows predeclared decision rules and is kept separate from the frozen test.

  \item On a frozen test set of $16$ previously unused MIT-BIH records, at the preselected setting $(m,\eps,\deltap)=(40,1,10^{-5})$, the structured release reduces median relative $L_2$ error by $39.6\%$ and frequency-band error by $51.8\%$ compared with the median of the independently calibrated random controls. It also outperforms each of the five calibrated random masks on both summary metrics. The preselected point is the qualifying development point with the strongest privacy and the smallest development margin in relative $L_2$, and the test margin exceeds the development margin.
\end{itemize}

At this setting, the comparisons show that retaining the low-order DCT modes improves utility beyond compression alone. A post-hoc development ablation (\cref{sec:ablation}) attributes most of this gain to the constant DCT mode, which carries the curve's overall level; smooth structure beyond that mode pays off at larger privacy budgets. The guarantee is window-level, not patient-level; we do not claim basis optimality or deployment performance.

\section{Private Spectral Statistics and Privacy Model}
\label{sec:problem}

\paragraph{Spectral statistic.}
Let $\windowVec_i\in\R^T$ be an $8$-second electrocardiogram (ECG) window; at $360\,$Hz, $T=2{,}880$ samples. The implementation divides the signal by a public $5\,$mV bound and clips it coordinatewise to $[-1,1]$. From $\windowVec_i$, we compute a Welch estimate~\citep{welch1967} of the power spectral density (PSD) and keep the bins in a fixed frequency interval. The per-window statistic is the log-PSD vector
\begin{equation}
  \statVec_i=\log_{10}\!\left(\max\{\PSD(\windowVec_i),\rho\}\right)\in\R^{\fullDim},
  \label{eq:logpsd}
\end{equation}
where $\rho=10^{-12}$ is a fixed numerical floor and the maximum is taken coordinatewise. For our Welch configuration, $\fullDim=142$ and the realized grid ranges from $0.703125$ to $99.84375\,$Hz.

For a group $D=(\statVec_1,\ldots,\statVec_{\groupSize})$, the non-private target is the mean curve
\begin{equation}
  \targetMean(D)=\frac{1}{\groupSize}\sum_{i=1}^{\groupSize}\statVec_i.
  \label{eq:target-mean}
\end{equation}
We release a summary of power over frequency, not the ECG waveform, and the logarithm compresses the wide range of power values. With the normalization above, the log-PSD values are large and negative, so a constant offset dominates the Euclidean norm of every curve. Fixed-duration windows provide comparable local spectral summaries, and their mean defines the aggregate query studied here. The method approximates this mean, not an individual ECG waveform. The averaging in \cref{eq:target-mean} also removes variation between windows, which may matter for downstream tasks.

\paragraph{Privacy model.}
We use standard $(\eps,\deltap)$-DP with the following adjacency relation.

\begin{definition}[Window-level replace-one adjacency]
\label[definition]{def:adjacency}
Two groups $D,D'\in(\R^{\fullDim})^{\groupSize}$ are adjacent, written $D\simeq D'$, if they have equal size and differ in the log-PSD vector of at most one ECG window.
\end{definition}

For $\eps\geq0$ and $0\leq\deltap<1$, a randomized mechanism $\cM$ is $(\eps,\deltap)$-DP if, for every pair of adjacent groups $D\simeq D'$ and every measurable event $A$, $\Pr[\cM(D)\in A]\leq e^{\eps}\Pr[\cM(D')\in A]+\deltap$.

The privacy unit is one bounded $8$-second window. A group contains $\groupSize$ windows from one record, and adjacency replaces at most one of them; thus \cref{thm:dp} gives window-level, not patient-level, privacy. Each released curve summarizes part of one person's recording ($40$ of its $120$ windows at $m=40$). The guarantee limits what the release reveals about any single window, such as a transient episode, but it is not designed to hide that person's typical spectrum. Releases over disjoint groups compose in parallel (\cref{app:disjoint}), while reusing a window or repeatedly releasing one group requires sequential composition.

\paragraph{Utility.}
For a released curve $\releasedStat$ and target $\targetMean=\targetMean(D)$, the relative $L_2$ error is
\begin{equation}
  \operatorname{RelL2}(\releasedStat,\targetMean)
  =\frac{\norm{\releasedStat-\targetMean}}{\max\{\norm{\targetMean},10^{-12}\}}.
  \label{eq:rel-l2}
\end{equation}
The denominator is dominated by the constant offset noted above, so relative $L_2$ error depends on the power units; band error, defined below, has no such normalization. Cosine similarity measures the alignment of the two curves. Because the curves are not mean-centered, it mainly reflects agreement in overall level. Let $\mathcal B_1,\ldots,\mathcal B_4$ denote the bins in $[0.1,5)$, $[5,15)$, $[15,40)$, and $[40,100]\,$Hz, with the last interval including the nominal upper endpoint. The realized Welch grid ends at $99.84375\,$Hz, so no sampled bin lies exactly at $100\,$Hz. The band error is
\begin{equation}
  \operatorname{BandErr}(\releasedStat,\targetMean)
  =\frac{1}{4}\sum_{b=1}^{4}
  \left|\frac{1}{|\mathcal B_b|}\sum_{j\in\mathcal B_b}\releasedStat_j
  -\frac{1}{|\mathcal B_b|}\sum_{j\in\mathcal B_b}\targetMean_j\right|.
  \label{eq:band-error}
\end{equation}
Because $\statVec_i$ is a $\log_{10}$ quantity, band error is measured in decades of power. For the two error measures, lower is better; for cosine similarity, higher is better.

\section{Structured Private Spectral Release}
\label{sec:mechanism}

Let $\dctMat\in\R^{\fullDim\times\fullDim}$ be an orthonormal type-II discrete cosine transform (DCT) matrix~\citep{ahmed1974dct}; hence $\dctMat^\top\dctMat=I_{\fullDim}$. Let $\selector\in\{0,1\}^{\projDim\times\fullDim}$ select the first $\projDim$ DCT coordinates. The projected representation of one log-PSD curve is
\begin{equation}
  \projVec_i=\selector\dctMat\statVec_i\in\R^{\projDim}.
\end{equation}
For $C>0$, define Euclidean clipping by
\begin{equation}
  \clip_C(v)=v\min\!\left\{1,\frac{C}{\norm{v}}\right\}.
  \label{eq:clip}
\end{equation}
The deterministic projected query is
\begin{equation}
  \query(D)=\frac{1}{\groupSize}\sum_{i=1}^{\groupSize}
  \clip_C(\selector\dctMat\statVec_i).
  \label{eq:query}
\end{equation}
The mechanism adds $\eta\sim\cN(0,\noiseStd^2I_{\projDim})$ and maps the result back to the original frequency grid:
\begin{align}
  \releasedVec&=\query(D)+\eta,\label{eq:noisy-proj}\\
  \releasedStat&=\dctMat^\top\selector^\top\releasedVec.
  \label{eq:reconstruct}
\end{align}
The multiplication by $\selector^\top$ zero-fills the omitted DCT coordinates.

\begin{algorithm}[H]
\caption{Structured private release of an aggregate spectral statistic}
\label{alg:release}
\begin{algorithmic}[1]
\Require Bounded windows $\windowVec_1,\ldots,\windowVec_m$; public $(\projDim,\clipRad)$; privacy parameters $(\eps,\deltap)$
\For{$i=1,\ldots,m$}
  \State Compute $\statVec_i$ by the fixed Welch and log transform in \cref{eq:logpsd}.
  \State $\projVec_i\gets\selector\dctMat\statVec_i$.
  \State $\clippedVec_i\gets\clip_{\clipRad}(\projVec_i)$ using \cref{eq:clip}.
\EndFor
\State $q\gets m^{-1}\sum_{i=1}^{m}\clippedVec_i$.
\State Set $\noiseStd\gets\noiseMult(2\clipRad/m)$ and sample $\eta\sim\cN(0,\noiseStd^2I_{\projDim})$.
\State $\releasedVec\gets q+\eta$.
\State \Return $\releasedStat\gets\dctMat^\top\selector^\top\releasedVec$.
\end{algorithmic}
\end{algorithm}

\subsection{Privacy guarantee}
\label{sec:privacy}
The amplitude clip and the floor $\rho$ bound every log-PSD coordinate, but the resulting sensitivity bound is very loose. We obtain a tight bound by clipping each projected vector before aggregation.

\begin{lemma}[Sensitivity]
\label[lemma]{lem:sensitivity}
For the query in \cref{eq:query} and window-level replace-one adjacency,
\begin{equation}
  \sup_{D\simeq D'}\norm{\query(D)-\query(D')}\le\frac{2C}{m}.
  \label{eq:sensitivity}
\end{equation}
\end{lemma}
\begin{proof}
Adjacent groups differ in one summand. Both possible clipped vectors lie in the closed $L_2$ ball of radius $C$, so their distance is at most $2C$. Division by $m$ gives \cref{eq:sensitivity}.
\end{proof}

Let $\noiseMult>0$ be any value satisfying
\begin{equation}
  \min_{\alpha>1}\left\{\frac{\alpha}{2\noiseMult^2}
  +\frac{\log(1/\deltap)}{\alpha-1}\right\}\le\eps,
  \label{eq:rdp-calibration}
\end{equation}
which is the R\'enyi-DP accounting condition for a single Gaussian release~\citep{mironov2017rdp}. This conversion is not tight for a single release; the analytic Gaussian mechanism~\citep{balle2018analytic} would permit a smaller multiplier (about $24\%$ smaller at $(\eps,\deltap)=(1,10^{-5})$) for every method alike. We keep the R\'enyi-DP calibration used in the frozen protocol.

\begin{theorem}[Window-level differential privacy]
\label{thm:dp}
Set $\noiseStd=\noiseMult(2C/m)$, where $\noiseMult$ satisfies \cref{eq:rdp-calibration}. Then \cref{alg:release} is $(\eps,\deltap)$-differentially private under window-level replace-one adjacency.
\end{theorem}
\begin{proof}
By \cref{lem:sensitivity}, the $L_2$ sensitivity of $\query$ is at most $2C/m$. The Gaussian mechanism with standard deviation $\noiseMult(2C/m)$ is therefore $(\alpha,\alpha/(2\noiseMult^2))$-R\'enyi DP for every $\alpha>1$, which implies $(\eps_\alpha,\deltap)$-DP with $\eps_\alpha=\alpha/(2\noiseMult^2)+\log(1/\deltap)/(\alpha-1)$~\citep{mironov2017rdp}. By \cref{eq:rdp-calibration}, the best such $\alpha$ gives $\eps_\alpha\le\eps$, so $\releasedVec$ is $(\eps,\deltap)$-DP. Zero-filling and the inverse DCT are deterministic post-processing.
\end{proof}
As is standard, the guarantee assumes exact real-valued Gaussian sampling. A floating-point implementation needs a discrete or otherwise secure sampler to retain it formally.

\subsection{Error and public calibration}
\label{sec:calibration}
Because $\dctMat^\top\selector^\top$ has orthonormal columns, the expected per-coordinate squared reconstruction error has the exact form
\begin{align}
  \frac{1}{d}\E\norm{\releasedStat-\targetMean(D)}^2
  &=\frac{1}{d}\norm{\dctMat^\top\selector^\top\query(D)-\targetMean(D)}^2
    +\frac{k}{d}\noiseStd^2,
  \label{eq:error-decomp}
\end{align}
where the expectation is over the Gaussian noise (derivation in \cref{app:identity}). The first term combines projection and clipping distortion. The second is the expected energy of $k$ independent Gaussian coordinates after orthonormal reconstruction. Substituting $\noiseStd=\noiseMult(2C/m)$ gives
\begin{equation}
  E_{\mathrm{noise}}(k,C)=\frac{k}{d}\left(\noiseMult\frac{2C}{m}\right)^2.
  \label{eq:noise-energy}
\end{equation}
When this term dominates, a release that keeps $k$ coordinates and a full-dimensional release with nearly the same radius differ in root-mean-square error by the factor $\sqrt{k/d}$. For $k=7$ and $d=142$, the factor is $0.222$.

The method chooses $(k,C)$ before seeing private or evaluation records. Let $\mathcal G_{\mathrm{pub}}$ be groups formed from public calibration windows. For candidate sets $\mathcal K$ and $\mathcal C_k$, we minimize
\begin{equation}
  (k^*,C^*)\in\argmin_{k\in\mathcal K,\,C\in\mathcal C_k}
  \left[\frac{1}{|\mathcal G_{\mathrm{pub}}|}\sum_{G\in\mathcal G_{\mathrm{pub}}}
  \frac{1}{d}\norm{\dctMat^\top P_k^\top q_{k,C}(G)-\targetMean(G)}^2
  +\frac{k}{d}\left(\noiseMult\frac{2C}{m}\right)^2\right].
  \label{eq:public-objective}
\end{equation}
The clipping candidates are public quantiles of projected norms. This selection is data-dependent only through records designated public; it is not charged to the privacy budget of the protected groups. The approach assumes that public records exist from a comparable population and recording setup. In our benchmark every MIT-BIH record is public, so the split between public and protected records is simulated.

\section{Experimental Protocol}
\label{sec:experiments}

\paragraph{Data and spectral representation.}
We use the MIT-BIH Arrhythmia Database, which contains $48$ half-hour excerpts of two-channel ambulatory electrocardiogram (ECG) recordings from $47$ subjects, sampled at $360\,$Hz~\citep{moody2001impact,mitbih,goldberger2000physionet}. The confirmatory protocol adopts the widely used $44$-record non-paced subset from the inter-patient protocol of \citet{dechazal2004}. The paced-beat records 102, 104, 107, and 217 were therefore outside public calibration, development, and frozen-test selection. This restriction, inherited from a classification protocol, is not a condition of the mechanism in \cref{sec:mechanism}, which is defined for any bounded window as in \cref{sec:problem}. After the frozen final test, we applied the unchanged operating point to these four records and, separately, to the complete $48$-record database. Those analyses are explicitly post-hoc and are reported in \cref{app:posthoc-paced,app:all48}; neither enters the confirmatory decision.

For the confirmatory protocol, we take the first channel of each included record \rev{(modified limb lead II, MLII, in all calibration and final-test records and in $43$ of the $44$ records overall; record 114, in the secondary development cohort, stores lead V5 first)}, form non-overlapping $8$-second windows, normalize by $5\,$mV, and clip to $[-1,1]$. Each record contributes $120$ windows\rev{, evenly spaced among its $225$ non-overlapping windows}, from which disjoint groups of \rev{$m$ consecutive windows} are formed within that record; no group combines windows from different records. Welch estimation uses a Hann window, $512$ samples per segment, $256$ samples of overlap, constant detrending, and density scaling. We keep the power spectral density (PSD) bins in $0.1$--$100\,$Hz; the resulting grid contains $142$ bins from $0.703125$ to $99.84375\,$Hz at $0.703125\,$Hz spacing. We apply \cref{eq:logpsd} to obtain the per-window statistic.

\paragraph{Separation of records.}
Within the $44$-record confirmatory universe, six records are used only for public calibration, six form the primary development cohort, and sixteen form a secondary development cohort. A separate set of sixteen records remains untouched until the final operating point has been fixed. The complete chronology and record assignment are listed in \cref{tab:record-splits}. No final-test record participates in parameter selection. Records 201 and 202 come from the same subject~\citep{mitbih}; record 201 is in the primary development cohort and record 202 in the frozen test, so records are separated by recording, not strictly by subject.

\paragraph{Development grid and final point.}
The development study evaluates
\[
  m\in\{10,20,40\},\qquad \eps\in\{0.5,1,2,4,8\},\qquad \deltap=10^{-5}.
\]
The public grid uses $k/d\in\{0.05,\allowbreak 0.10,\allowbreak 0.20,\allowbreak 0.40,\allowbreak 0.60,\allowbreak 0.80,\allowbreak 1\}$ and clipping quantiles $\{0.50,\allowbreak 0.70,\allowbreak 0.85,\allowbreak 0.95\}$; \rev{each ratio is multiplied by $d$ and rounded to the nearest integer, so} the ratio $0.05$ gives $k=7$. The same objective in \cref{eq:public-objective} is used throughout. Public calibration selects $k=7$ at every development point. A point qualifies under the predeclared rule if, compared with the median of the five calibrated random masks, the structured release reduces median relative $L_2$ and median band error by at least $5\%$ each; reduces median relative $L_2$ by at least $5\%$ compared with full-dimensional release (Full DP, defined below); and wins jointly (lower relative $L_2$ and lower band error) against at least four of the five masks. The same rule decides the final test. After the development study with the independently calibrated random controls\rev{, which are a stronger baseline than the matched random control defined below} (\cref{sec:results}), the qualifying point with the strongest privacy is frozen as
\begin{equation}
  (m,\eps,\deltap,k)=(40,1,10^{-5},7).
  \label{eq:final-point}
\end{equation}
The final test runs only \cref{eq:final-point}; it performs no optimization or recalibration.

\paragraph{Baselines.}
We compare four families. All except Full DP use the orthonormal discrete cosine transform (DCT) of \cref{sec:mechanism}.
\begin{enumerate}[leftmargin=1.5em]
  \item \textbf{Full DP}, a full-dimensional differentially private release, clips and perturbs all $d=142$ log-PSD coordinates. \rev{Its clipping radius is selected with the same public objective and quantile grid.}
  \item \textbf{Matched random DCT} keeps a fixed random set of $k$ DCT coordinates and uses the same $k$, $C$, noise multiplier, groups, and paired standard-normal draw as the structured prefix.
  \item \textbf{Calibrated random DCT} uses five fixed masks, generated with seeds 819--823. Each mask keeps the same $k$ but selects its own clipping radius from the same public quantile grid and objective.
  \item \textbf{Structured prefix} is \cref{alg:release}, which retains the first $k$ DCT modes of the ordered log-PSD curve.
\end{enumerate}
The calibrated random controls separate the choice of coordinates from their number. A win over Full DP alone could be explained by dimension reduction, while a win over independently calibrated random subspaces of the same size shows that the choice of coordinates matters. A uniformly drawn mask of $7$ of the $142$ coordinates omits the constant DCT mode with probability $135/142\approx0.95$, and none of our five masks contains it. These controls therefore cannot isolate the constant mode from the other low-order modes; \cref{sec:ablation} addresses that question. Random projection has a broader history in private dimensionality reduction~\citep{blocki2012jl,gondara2020dprp}; here we hold the released dimension fixed and vary which transform coordinates are retained.

\paragraph{Reporting and reproducibility.}
For each randomized method, three independent Gaussian draws are used as Monte Carlo evaluation replicates. They represent separate possible releases, not three outputs jointly covered by one privacy budget. We report medians across groups and replicates; means and standard deviations are provided in the released artifact, and in \cref{tab:final-detail} for the final test. The code snapshot and frozen configurations are dated \codefreeze{} and are available in the GitHub repository.\footnote{\url{https://github.com/onbasligroup/structure-matters}} Experiments were run on Windows 11 Pro with an Intel Core i9-14900HX and $32\,\mathrm{GB}$ of RAM, using CPU-only Python~3.12.13. The repository contains the frozen configurations, raw and summary results, and scripts used to reproduce the reported tables and figures.

\section{Results}
\label{sec:results}

\subsection{Development with independently calibrated random controls}
\label{sec:dev-results}
Public calibration selects the first $k=7$ discrete cosine transform (DCT) modes, i.e., $4.93\%$ of the $142$ log-PSD coordinates, at all fifteen development points. It selects $C=74.7546$ at fourteen points and $C=76.2370$ only at $(m,\eps)=(40,8)$. In contrast, the independently calibrated random masks select radii between $0.3639$ and $5.2269$. Thus, the random baselines are not disadvantaged by being forced to carry the structured method's much larger sensitivity.

\begin{figure}[t]
  \centering
  \includegraphics[width=\textwidth]{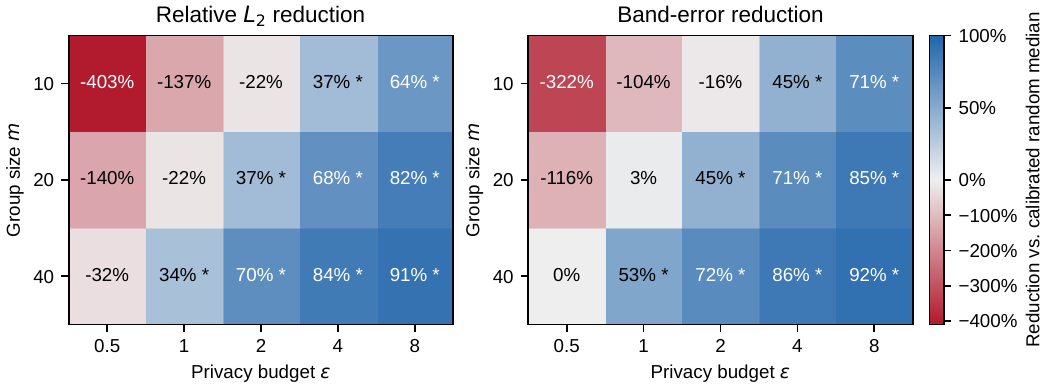}
  \caption{Development reduction relative to the median of five independently calibrated random DCT masks (left: relative $L_2$ error; right: band error). Positive values favor the structured prefix. An asterisk marks a point that satisfies the complete predeclared rule of \cref{sec:experiments}: at least $5\%$ reduction in both metrics relative to the random median, at least $5\%$ relative-$L_2$ reduction relative to full-dimensional differentially private (DP) release (Full DP), and joint wins against at least four of five masks. Nine of fifteen points qualify.}
  \label{fig:dev-heatmap}
\end{figure}

\Cref{fig:dev-heatmap} shows the two primary margins. Nine of fifteen $(m,\eps)$ pairs satisfy the frozen rule: $(10,4)$, $(10,8)$; $(20,2)$, $(20,4)$, $(20,8)$; and $(40,1)$, $(40,2)$, $(40,4)$, $(40,8)$. At every qualifying point, the structured prefix beats all five random masks in both relative $L_2$ and band error (a \emph{joint win} over each mask). The same pattern appears on the secondary development cohort, which is not used in the decision\rev{: the same nine points qualify there, and at $(40,1)$ the reductions are $39.2\%$ in relative $L_2$ and $51.3\%$ in band error, with joint wins against all five masks}. The full development table is in \cref{app:development}.

The low-budget boundary is informative. The six non-qualifying points form the low-budget corner of the grid: $m=10$ with $\eps\le2$, $m=20$ with $\eps\le1$, and $m=40$ with $\eps=0.5$. At all six, the structured release has higher median relative $L_2$ than the median independently calibrated random control, and at four of them it also has higher band error. Increasing $m$ reduces sensitivity $2C/m$, while increasing $\eps$ reduces the required noise multiplier. Consistent with the Gaussian term in \cref{eq:noise-energy}, the structured prefix becomes favorable in both metrics as the noise contribution decreases across the development grid.

The gain over Full DP, in contrast, is nearly constant. It lies between $77.4\%$ and $79.8\%$ at all fifteen development points (\cref{tab:development}). The noise term in \cref{eq:noise-energy} predicts this value: with nearly equal radii, the ratio of root-mean-square errors is $\sqrt{k/d}=\sqrt{7/142}=0.222$, a reduction of $77.8\%$.

\subsection{Frozen untouched test}
\label{sec:final-results}
The final point in \cref{eq:final-point} was selected before any final-test record was downloaded. The final cohort contains $16$ records from the predefined $44$-record confirmatory universe, $1{,}920$ windows, and $48$ disjoint groups of size $40$. The structured release uses $C=74.754581$; Full DP uses $C=74.785476$; and the five calibrated random masks use the frozen radii $2.59736$, $0.363866$, $0.452277$, $0.479546$, and $0.44379$ (\cref{tab:random-masks}). This is the sole confirmatory evaluation used by the frozen decision rule.

\begin{table}[t]
\centering
\caption{Untouched final-test medians at $(m,\eps,\deltap)=(40,1,10^{-5})$. The calibrated-random row is the median across five masks, each calibrated independently on public records. \rev{For reference, releasing the zero vector gives a relative $L_2$ of exactly $1$.}}
\label{tab:final}
\small
\begin{tabular}{lccc}
\toprule
Method & Relative $L_2$ $\downarrow$ & Band error $\downarrow$ & Cosine $\uparrow$ \\
\midrule
Full DP & 2.805 & 3.300 & 0.328 \\
Matched random DCT & 1.165 & 5.205 & 0.002 \\
Calibrated random DCT (median) & 1.000 & 5.011 & 0.004 \\
\textbf{Structured prefix} & \textbf{0.604} & \textbf{2.415} & \textbf{0.866} \\
\bottomrule
\end{tabular}
\end{table}

\Cref{tab:final} gives the frozen result. Relative to the median calibrated-random control, the structured release reduces median relative $L_2$ by
\[
  1-\frac{0.604407}{0.999993}=39.56\%,
\]
and reduces band error by
\[
  1-\frac{2.415225}{5.010914}=51.80\%.
\]
\rev{A record-level bootstrap of the stored per-group results ($10{,}000$ resamples of the $16$ records, computed after the confirmatory run) gives $95\%$ percentile intervals of $35\%$ to $41\%$ for the relative-$L_2$ reduction and $45\%$ to $58\%$ for the band-error reduction.} The structured release also reduces relative $L_2$ by $78.45\%$ compared with Full DP and wins jointly against all five calibrated random masks. Compared with the matched random DCT control, which uses the structured radius $C$, the reductions are $48.12\%$ in relative $L_2$ and $53.60\%$ in band error. These margins reproduce the development ordering without choosing among test points. The frozen point had the smallest relative-$L_2$ margin of the nine qualifying development points ($34.3\%$), and the test margin is larger, so selecting this point did not inflate the expected gain. The $78.45\%$ reduction relative to Full DP matches the $77.8\%$ predicted by \cref{eq:noise-energy}.

The number of noisy coordinates alone does not explain the result. The calibrated random mechanisms retain the same number of coordinates, $k=7$, and use much smaller clipping radii, yet their median cosine similarity is $0.0038$, against $0.866$ for the structured prefix. All five masks have median relative $L_2$ error within $0.1\%$ of $1$, the value obtained by releasing the zero vector. Without the constant mode, the seven random coefficients carry little of the curve's energy. Public calibration selects small radii for these masks, and the resulting releases convey almost nothing about the target.

Most of the contrast therefore comes from a single coefficient. The DCT prefix contains the constant mode, which carries the overall level of the log-PSD curve, and none of the five random masks does. \Cref{sec:ablation} isolates its contribution. In absolute terms the release at this privacy level is coarse: a median band error of $2.4$ decades means that band powers are off by \rev{more than} two orders of magnitude, and the released curve mainly conveys the overall level of the target. \rev{The Pearson correlation between released and target curves, which ignores the level, is correspondingly lower than the cosine similarity: its median is $0.41$ for the structured prefix, $0.07$ for Full DP, and $0.02$ for the median calibrated random mask.}

\subsection{Post-hoc fixed-policy robustness}
\label{sec:posthoc-results}
After the frozen decision was recorded, we applied the unchanged mechanism and operating point to two supplementary settings. On the four paced-beat records, the structured release reduces median relative $L_2$ by $35.5\%$ and band error by $53.2\%$ relative to the median of the five calibrated random masks, with joint wins against all five. A retrospective pooled analysis over all $48$ records gives reductions of $38.6\%$ and $46.3\%$, respectively, again with five joint wins. These are not additional final tests: the paced-record analysis was specified after the confirmatory run, and every record in the all-$48$ pool already had a calibration, development, confirmatory, or robustness role. The radius was not recalibrated, although the clipping rate fell from $88.0\%$ on the final cohort to $3.1\%$ on the paced records. The complete per-record results, the clipping-rate shift on paced signals, and the fixed-policy all-database analysis are reported in \cref{app:posthoc-paced,app:all48}.

\subsection{Post-hoc constant-mode ablation}
\label{sec:ablation}
None of the five random masks contains the constant DCT mode, which carries the overall level of the log-PSD curve and a median of about $94$--$95\%$ of a window's energy. After the frozen test, we therefore ran a development-only ablation with five \emph{DC-forced} random masks: each original mask with one index replaced by the constant mode, and with its own public clipping radius. A point qualifies if, compared with the median of these masks, the structured prefix reduces median relative $L_2$, median band error, and median mean-centered relative $L_2$ by at least $5\%$ each, and wins jointly against at least four of the five. This rule was fixed before the run.

The structured prefix qualifies at six of the fifteen development points, all at larger budgets: $(m,\eps)=(10,8)$, $(20,4)$, $(20,8)$, $(40,2)$, $(40,4)$, and $(40,8)$. It does not qualify at the frozen point $(40,1)$, where it reduces relative $L_2$ by only $1.8\%$ and has higher band error than the DC-forced controls. At that point, the gain over the original random controls is therefore attributable mainly to the constant mode. Releasing the constant mode alone ($k=1$) gives lower relative $L_2$ than $k=7$ at twelve of the fifteen points, and public calibration over a grid extended to $k\in\{1,\ldots,6\}$ selects $k\le3$ throughout. Details are in \cref{app:ablation}. The frozen $k=7$ result therefore understates what a better-calibrated prefix can achieve. These findings are post-hoc and have not been confirmed on untouched data.

\section{Related Work}
\label{sec:related}

\paragraph{Transformation-based release.}
\citet{rastogi2010fpa} introduced the Fourier Perturbation Algorithm for differentially private (DP) aggregate time series. It retains a limited number of Fourier coefficients, adds Laplace noise, and reconstructs a time-domain query answer under pure DP. Our released object is instead an aggregate log-power spectral density (log-PSD) curve, and the mechanism uses projected $L_2$ clipping with Gaussian $(\eps,\deltap)$-DP. \citet{parker2022spdp} define spectral differential privacy for the frequency content of wide-sense stationary processes and study smart-meter data. Their privacy notion and stochastic-process objective differ from standard replace-one privacy for a finite collection of signal windows.

\paragraph{ECG privacy and spectral learning.}
\citet{bonomi2022ecg} sanitize individual electrocardiogram (ECG) segments using a compact discrete cosine transform (DCT) representation and metric privacy, then evaluate aggregate analytics. This is the closest ECG-specific use of DCT, but it protects an individual waveform under a metric-based indistinguishability definition, whereas we release a central aggregate statistic under replace-one $(\eps,\deltap)$-DP. Spectral-DP perturbs and filters neural-network gradients in a transform domain~\citep{feng2023spectraldp}. Its protected object is a model update, not a spectral summary of measured signals. We therefore do not call our method spectral DP, and we do not claim the first use of spectral transforms in DP.

\paragraph{Private low-dimensional representations.}
Random projections can support private dimensionality reduction and release~\citep{blocki2012jl,gondara2020dprp}. Our random controls serve a narrower purpose. They keep exactly the same number of DCT coordinates as the structured prefix, and five masks receive independent public clipping calibration. The comparison asks whether ordering on a physical frequency curve provides utility beyond an equally sized generic transform subspace.

\section{Limitations, Future Work, and Conclusion}
\label{sec:conclusion}

The mechanism gives window-level replace-one differential privacy (DP): one bounded electrocardiogram (ECG) window is the protected unit, not all windows contributed by one person. It therefore does not provide patient-level DP, and repeated releases involving the same window require composition. A stronger extension would define person-level adjacency over all windows contributed by one participant and bound that participant's total contribution, requiring a new sensitivity and composition analysis.

The evaluation uses one ECG database and one statistic family, aggregate
Welch log-power spectral density (log-PSD) curves. The public calibration records come from the same database, lead, and recording setup as the protected records. The reported errors measure fidelity to the
non-private mean, not diagnostic accuracy; whether lower spectral error
improves downstream classification remains untested. At the frozen setting the absolute error is still large, with a median band error of $2.4$ decades. \rev{A non-private curve averaged from the public calibration records is expected to lie closer to each group mean than any private release compared here. We did not evaluate that baseline; our comparison isolates the choice of coordinates at a fixed privacy level.} Ongoing work is
extending the study to larger and more heterogeneous ECG datasets with
subject-disjoint partitions. Generalization to other time-series domains
is not established. For wearable or edge deployment, the private output
can be represented by $k$ noisy coefficients and reconstructed at the
receiver, but communication savings and device-side computation, memory,
energy, and latency are not measured here.

Public records are required for calibration. The search selects the smallest candidate dimension, $k=7$, at every point, while many random controls select the lowest clipping quantile. Both choices lie on the edge of the grid. We therefore do not claim global optimality over bases, dimensions, clipping thresholds, or random projections. The post-hoc ablation in \cref{sec:ablation} indicates that $k=7$ is larger than needed, and that the gain at the frozen point is attributable mainly to the constant discrete cosine transform (DCT) mode; these development-only findings are not confirmed on untouched data. Comparisons with wavelets and PCA remain future work. The paced-record and all-$48$ analyses are post-hoc robustness checks and do not alter the frozen confirmatory decision.

At the preselected evaluation setting, projecting the log-PSD curve onto
low-order DCT modes before clipping and Gaussian perturbation yields
lower relative $L_2$ and frequency-band error than full-dimensional
release and each of the five independently calibrated random DCT
subspaces. Most of this advantage comes from retaining the curve's overall level, the constant DCT mode.
These results support matching the retained subspace to the
statistic, but do not establish a uniform advantage across privacy
budgets, group sizes, or signal domains.

\section*{Acknowledgments}
Reza Saadi thanks Prof.\ Alptekin Küpçü for the opportunity to work in the KU-Crypto Lab.

\bibliography{refs}

\begin{thebibliography}{15}
\providecommand{\natexlab}[1]{#1}
\providecommand{\url}[1]{\texttt{#1}}
\expandafter\ifx\csname urlstyle\endcsname\relax
  \providecommand{\doi}[1]{doi: #1}\else
  \providecommand{\doi}{doi: \begingroup \urlstyle{rm}\Url}\fi

\bibitem[Dwork and Roth(2014)]{dworkroth2014}
Cynthia Dwork and Aaron Roth.
\newblock The algorithmic foundations of differential privacy.
\newblock \emph{Foundations and Trends in Theoretical Computer Science},
  9\penalty0 (3--4):\penalty0 211--407, 2014.
\newblock \doi{10.1561/0400000042}.

\bibitem[Rastogi and Nath(2010)]{rastogi2010fpa}
Vibhor Rastogi and Suman Nath.
\newblock Differentially private aggregation of distributed time-series with
  transformation and encryption.
\newblock In \emph{Proceedings of the 2010 ACM SIGMOD International Conference
  on Management of Data}, pages 735--746. ACM, 2010.
\newblock \doi{10.1145/1807167.1807247}.

\bibitem[Parker et~al.(2022)Parker, Hale, and Barooah]{parker2022spdp}
Kendall Parker, Matthew Hale, and Prabir Barooah.
\newblock Spectral differential privacy: Application to smart meter data.
\newblock \emph{IEEE Internet of Things Journal}, 9\penalty0 (7):\penalty0
  4987--4996, 2022.
\newblock \doi{10.1109/JIOT.2021.3107770}.

\bibitem[Bonomi et~al.(2022)Bonomi, Wu, and Fan]{bonomi2022ecg}
Luca Bonomi, Zeyun Wu, and Liyue Fan.
\newblock Sharing personal {ECG} time-series data privately.
\newblock \emph{Journal of the American Medical Informatics Association},
  29\penalty0 (7):\penalty0 1152--1160, 2022.
\newblock \doi{10.1093/jamia/ocac047}.

\bibitem[Feng et~al.(2023)Feng, Xu, Wen, Venkitasubramaniam, and
  Ding]{feng2023spectraldp}
Ce~Feng, Nuo Xu, Wujie Wen, Parv Venkitasubramaniam, and Caiwen Ding.
\newblock Spectral-{DP}: Differentially private deep learning through spectral
  perturbation and filtering.
\newblock In \emph{2023 IEEE Symposium on Security and Privacy (SP)}, pages
  1944--1960, 2023.
\newblock \doi{10.1109/SP46215.2023.10179457}.

\bibitem[Welch(1967)]{welch1967}
Peter~D. Welch.
\newblock The use of fast {Fourier} transform for the estimation of power
  spectra: A method based on time averaging over short, modified periodograms.
\newblock \emph{IEEE Transactions on Audio and Electroacoustics}, 15\penalty0
  (2):\penalty0 70--73, 1967.
\newblock \doi{10.1109/TAU.1967.1161901}.

\bibitem[Ahmed et~al.(1974)Ahmed, Natarajan, and Rao]{ahmed1974dct}
Nasir Ahmed, T.~Raj Natarajan, and K.~R. Rao.
\newblock Discrete cosine transform.
\newblock \emph{IEEE Transactions on Computers}, C-23\penalty0 (1):\penalty0
  90--93, 1974.
\newblock \doi{10.1109/T-C.1974.223784}.

\bibitem[Mironov(2017)]{mironov2017rdp}
Ilya Mironov.
\newblock R{\'e}nyi differential privacy.
\newblock In \emph{2017 IEEE 30th Computer Security Foundations Symposium
  (CSF)}, pages 263--275, 2017.
\newblock \doi{10.1109/CSF.2017.11}.

\bibitem[Balle and Wang(2018)]{balle2018analytic}
Borja Balle and Yu-Xiang Wang.
\newblock Improving the {Gaussian} mechanism for differential privacy:
  Analytical calibration and optimal denoising.
\newblock In \emph{Proceedings of the 35th International Conference on Machine
  Learning (ICML)}, volume~80 of \emph{Proceedings of Machine Learning
  Research}, pages 394--403. PMLR, 2018.
\newblock URL \url{https://proceedings.mlr.press/v80/balle18a.html}.

\bibitem[Moody and Mark(2001)]{moody2001impact}
George~B. Moody and Roger~G. Mark.
\newblock The impact of the {MIT-BIH Arrhythmia Database}.
\newblock \emph{IEEE Engineering in Medicine and Biology Magazine}, 20\penalty0
  (3):\penalty0 45--50, 2001.
\newblock \doi{10.1109/51.932724}.

\bibitem[Moody and Mark(2005)]{mitbih}
George~B. Moody and Roger~G. Mark.
\newblock {MIT-BIH Arrhythmia Database} (version 1.0.0).
\newblock PhysioNet, 2005.
\newblock URL \url{https://doi.org/10.13026/C2F305}.

\bibitem[Goldberger et~al.(2000)Goldberger, Amaral, Glass, Hausdorff, Ivanov,
  Mark, Mietus, Moody, Peng, and Stanley]{goldberger2000physionet}
Ary~L. Goldberger, Luis A.~N. Amaral, Leon Glass, Jeffrey~M. Hausdorff,
  Plamen~Ch. Ivanov, Roger~G. Mark, Joseph~E. Mietus, George~B. Moody,
  Chung-Kang Peng, and H.~Eugene Stanley.
\newblock {PhysioBank}, {PhysioToolkit}, and {PhysioNet}: Components of a new
  research resource for complex physiologic signals.
\newblock \emph{Circulation}, 101\penalty0 (23):\penalty0 e215--e220, 2000.
\newblock \doi{10.1161/01.CIR.101.23.e215}.

\bibitem[de~Chazal et~al.(2004)de~Chazal, O'Dwyer, and Reilly]{dechazal2004}
Philip de~Chazal, Maria O'Dwyer, and Richard~B. Reilly.
\newblock Automatic classification of heartbeats using {ECG} morphology and
  heartbeat interval features.
\newblock \emph{IEEE Transactions on Biomedical Engineering}, 51\penalty0
  (7):\penalty0 1196--1206, 2004.
\newblock \doi{10.1109/TBME.2004.827359}.

\bibitem[Blocki et~al.(2012)Blocki, Blum, Datta, and Sheffet]{blocki2012jl}
Jeremiah Blocki, Avrim Blum, Anupam Datta, and Or~Sheffet.
\newblock The {Johnson--Lindenstrauss} transform itself preserves differential
  privacy.
\newblock In \emph{2012 IEEE 53rd Annual Symposium on Foundations of Computer
  Science}, pages 410--419, 2012.
\newblock \doi{10.1109/FOCS.2012.67}.

\bibitem[Gondara and Wang(2020)]{gondara2020dprp}
Lovedeep Gondara and Ke~Wang.
\newblock Differentially private small dataset release using random
  projections.
\newblock In \emph{Proceedings of the 36th Conference on Uncertainty in
  Artificial Intelligence (UAI)}, volume 124 of \emph{Proceedings of Machine
  Learning Research}, pages 639--648. PMLR, 2020.
\newblock URL \url{https://proceedings.mlr.press/v124/gondara20a.html}.

\end{thebibliography}

\appendix
\section{Additional Theory}
\label{app:theory}

\subsection{Exact reconstruction-error identity}
\label{app:identity}
Let $A=\dctMat^\top P_k^\top\in\R^{d\times k}$. Since $\dctMat$ is orthonormal and $P_kP_k^\top=I_k$, $A^\top A=I_k$. Write $a=Aq_{k,C}(D)-\targetMean(D)$ and $\eta\sim\cN(0,\tau^2I_k)$. Then
\begin{align}
\E\norm{a+A\eta}^2
&=\norm{a}^2+2\E\langle a,A\eta\rangle+\E\norm{A\eta}^2\\
&=\norm{a}^2+\E\eta^\top A^\top A\eta
=\norm{a}^2+k\tau^2.
\end{align}
The cross term is zero because $a$ is fixed given $D$ and $\E\eta=0$. Dividing by $d$ gives \cref{eq:error-decomp}.

\subsection{Disjoint groups}
\label{app:disjoint}
For a fixed $m$, each window is assigned to at most one group. If one window changes, at most one group output distribution changes. Parallel composition therefore gives the same window-level $(\eps,\deltap)$ guarantee for the vector of all disjoint group releases. This argument requires group membership to be fixed by window position, independently of window content\rev{, as in our protocol: each group consists of consecutive windows of one record}. It does not apply if windows are reused across groups or if several independent draws for one group are published.

\section{Extended Experimental Results}
\label{app:extended}

\subsection{Record partition and analysis chronology}
\label{app:record-splits}
\begin{table}[H]
\centering
\caption{Record-level chronology. The first four rows define the $44$-record confirmatory protocol. The paced-beat records were evaluated only after the frozen final decision.}
\label{tab:record-splits}
\small
\begin{tabular}{p{0.24\textwidth}p{0.68\textwidth}}
\toprule
Role & MIT-BIH record identifiers \\
\midrule
Public calibration & 100, 108, 116, 124, 208, 209 \\
Primary development & 113, 118, 119, 201, 207, 220 \\
Secondary development & 101, 103, 105, 106, 109, 111, 112, 114, 115, 117, 122, 203, 205, 215, 223, 230 \\
Frozen untouched test & 121, 123, 200, 202, 210, 212, 213, 214, 219, 221, 222, 228, 231, 232, 233, 234 \\
Paced-record post-hoc & 102, 104, 107, 217 \\
\bottomrule
\end{tabular}
\end{table}

The confirmatory record universe inherits the $44$-record non-paced convention of \citet{dechazal2004}. Thus, records 102, 104, 107, and 217 were outside calibration, development, and frozen-test selection. The restriction is procedural; \cref{alg:release} does not require it. Only after the confirmatory decision was recorded did we apply the unchanged mechanism to those four records and retrospectively to the complete database. Neither analysis can be merged into or used to revise the frozen decision.

Each record contributes $120$ non-overlapping windows\rev{, evenly spaced among the $225$ available}. Public calibration and primary development therefore use $720$ windows each; secondary development and the frozen final cohort use $1{,}920$ windows each. At $m=40$, the final cohort forms $48$ disjoint groups.

\subsection{Development margins and frozen final-test detail}
\label{app:development}
\begin{table}[H]
\centering
\caption{Development reductions. Columns three and four are relative to the median of five independently calibrated random masks; column five is relative to Full DP, the full-dimensional differentially private (DP) release. Entries are fractional reductions (e.g., $0.343$ is a $34.3\%$ reduction); positive values favor the structured prefix.}
\label{tab:development}
\small
\begin{tabular}{rrrrrrc}
\toprule
$m$ & $\eps$ & Rel-$L_2$ vs. random & Band vs. random & Rel-$L_2$ vs. Full & Joint wins & Qualifies \\
\midrule
10&0.5&$-$4.030&$-$3.220&0.780&0/5&No\\
10&1&$-$1.369&$-$1.043&0.796&0/5&No\\
10&2&$-$0.218&$-$0.157&0.794&0/5&No\\
10&4&0.370&0.453&0.798&5/5&Yes\\
10&8&0.640&0.706&0.780&5/5&Yes\\
20&0.5&$-$1.402&$-$1.165&0.794&0/5&No\\
20&1&$-$0.215&0.027&0.790&0/5&No\\
20&2&0.371&0.447&0.787&5/5&Yes\\
20&4&0.684&0.714&0.793&5/5&Yes\\
20&8&0.822&0.852&0.783&5/5&Yes\\
40&0.5&$-$0.317&0.001&0.775&0/5&No\\
40&1&0.343&0.531&0.774&5/5&Yes\\
40&2&0.702&0.720&0.798&5/5&Yes\\
40&4&0.836&0.857&0.790&5/5&Yes\\
40&8&0.906&0.918&0.774&5/5&Yes\\
\bottomrule
\end{tabular}
\end{table}
The frozen rule with the independently calibrated random controls\rev{, a stronger baseline than the matched random control,} passes at nine points. Selecting the smallest $\eps$ among qualifying points gives the frozen final point $(m,\eps)=(40,1)$.

\begin{table}[H]
\centering
\caption{Untouched final-test mean, median, and standard deviation over $48$ groups and three Monte Carlo draws.}
\label{tab:final-detail}
\small
\begin{tabular}{lrrr}
\toprule
Method and metric & Mean & Median & Std. dev. \\
\midrule
Structured relative $L_2$ & 0.6123 & 0.6044 & 0.1666 \\
Structured band error & 2.6614 & 2.4152 & 1.2502 \\
Structured cosine & 0.8447 & 0.8665 & 0.0881 \\
Full-DP relative $L_2$ & 2.8035 & 2.8051 & 0.1855 \\
Full-DP band error & 3.4011 & 3.3000 & 1.3506 \\
Full-DP cosine & 0.3247 & 0.3279 & 0.0727 \\
Matched-random relative $L_2$ & 1.1782 & 1.1650 & 0.0909 \\
Matched-random band error & 5.2563 & 5.2049 & 0.8719 \\
Matched-random cosine & 0.0032 & 0.0023 & 0.0198 \\
\bottomrule
\end{tabular}
\end{table}

\begin{table}[H]
\centering
\caption{Frozen calibrated-random controls on the untouched final test.}
\label{tab:random-masks}
\small
\begin{tabular}{rrrrrc}
\toprule
Mask seed & $C$ & Relative $L_2$ & Band error & Cosine & Joint win for structured \\
\midrule
819 & 2.597360 & 0.999062 & 5.122572 & 0.044009 & Yes \\
820 & 0.363866 & 0.999999 & 5.010914 & 0.002268 & Yes \\
821 & 0.452277 & 0.999993 & 5.010912 & 0.003844 & Yes \\
822 & 0.479546 & 0.999996 & 5.010914 & 0.003151 & Yes \\
823 & 0.443790 & 0.999989 & 5.008495 & 0.004834 & Yes \\
\bottomrule
\end{tabular}
\end{table}

\subsection{Post-hoc robustness on paced-beat records}
\label{app:posthoc-paced}
The four paced-beat records were evaluated once with the frozen point in \cref{eq:final-point}; no parameter was recalibrated. \rev{In records 102 and 104, the first stored channel is lead V5 instead of modified limb lead II.} The analysis contains $480$ windows, $12$ disjoint groups, and three Monte Carlo draws per group. \Cref{tab:robustness-summary} separates this post-hoc result from the confirmatory row, and \cref{tab:paced-records} gives the per-record results.

\begin{table}[H]
\centering
\caption{Confirmatory and post-hoc fixed-policy analyses. The first row is the sole confirmatory evaluation; the other rows are supplementary and do not enter the frozen rule. Reductions are relative to the median of five independently calibrated random masks, except for the Full-DP column.}
\label{tab:robustness-summary}
\small
\resizebox{\ifdim\width>\textwidth\textwidth\else\width\fi}{!}{%
\begin{tabular}{llrrrrrrr}
\toprule
Analysis & Status & Records & Rel-$L_2$ & Band & Cosine & Rel-$L_2$ red. & Band red. & Rel-$L_2$ red. vs. Full \\
\midrule
Frozen final & Confirmatory & 16 & 0.6044 & 2.4152 & 0.8665 & 39.56\% & 51.80\% & 78.45\% \\
Paced records & Post-hoc & 4 & 0.6451 & 2.2149 & 0.8634 & 35.49\% & 53.21\% & 79.23\% \\
All 48 records & Retrospective & 48 & 0.6143 & 2.6613 & 0.8655 & 38.57\% & 46.26\% & 78.38\% \\
\bottomrule
\end{tabular}}
\end{table}
All three analyses give joint wins against all five calibrated random masks.

\begin{table}[H]
\centering
\caption{Per-record paced-beat robustness under the frozen mechanism. Reductions are relative to the calibrated-random median.}
\label{tab:paced-records}
\small
\resizebox{\ifdim\width>\textwidth\textwidth\else\width\fi}{!}{%
\begin{tabular}{rrrrrrrr}
\toprule
Record & Rel-$L_2$ & Band & Cosine & Rel-$L_2$ red. & Band red. & Rel-$L_2$ red. vs. Full & Joint wins \\
\midrule
102 & 0.6101 & 2.1235 & 0.8477 & 38.99\% & 58.85\% & 79.68\% & 5/5 \\
104 & 0.7490 & 2.7317 & 0.7950 & 25.10\% & 44.07\% & 76.31\% & 5/5 \\
107 & 0.5491 & 1.5115 & 0.8896 & 45.09\% & 65.61\% & 82.26\% & 5/5 \\
217 & 0.4530 & 1.8251 & 0.9283 & 54.70\% & 60.69\% & 85.32\% & 5/5 \\
\bottomrule
\end{tabular}}
\end{table}

The paced subset has a mean structured pre-clip norm of $70.79$ and a mean clipping rate of $3.125\%$. The corresponding clipping rate is $68.385\%$ in the retrospective all-$48$ pool and $88.021\%$ in the original final cohort. This shift reflects lower projected norms under the unchanged radius $C=74.754581$; it is not the result of recalibration. The structured advantage survives the shift, including on record 104, the weakest paced case by relative $L_2$.

\subsection{Retrospective fixed-policy evaluation on all 48 records}
\label{app:all48}
The all-database analysis applies the same fixed mechanism to $5{,}760$ windows and $144$ disjoint groups. It is not an independent test: every record already had a role, $44$ in calibration, development, or the confirmatory test and the four paced records in the post-hoc robustness analysis. It is included to show how the fixed policy behaves on the complete database.

\begin{table}[H]
\centering
\caption{Retrospective comparison of the non-paced $44$-record pool, the paced four under the all-$48$ seed convention, and the complete database. Reductions are relative to the calibrated-random median.}
\label{tab:all48}
\small
\resizebox{\ifdim\width>\textwidth\textwidth\else\width\fi}{!}{%
\begin{tabular}{lrrrrrrr}
\toprule
Pool & Rel-$L_2$ & Band & Cosine & Rel-$L_2$ red. & Band red. & Rel-$L_2$ red. vs. Full & Joint wins \\
\midrule
Non-paced 44 & 0.6084 & 2.6746 & 0.8683 & 39.16\% & 46.31\% & 78.51\% & 5/5 \\
Paced four & 0.6737 & 2.5061 & 0.8411 & 32.63\% & 47.07\% & 77.91\% & 5/5 \\
All 48 & 0.6143 & 2.6613 & 0.8655 & 38.57\% & 46.26\% & 78.38\% & 5/5 \\
\bottomrule
\end{tabular}}
\end{table}

The paced-only result in \cref{tab:robustness-summary} and the paced subgroup in \cref{tab:all48} use different deterministic Monte Carlo streams because their group indices differ. The changes in relative-$L_2$ reduction ($35.49\%$ versus $32.63\%$) and band-error reduction ($53.21\%$ versus $47.07\%$) do not change the ordering or the $5/5$ joint wins. As a code-integrity check, replaying the original final cohort reproduced the archived raw metrics to a maximum absolute difference of $5.68\times10^{-14}$.

\subsection{Post-hoc constant-mode ablation}
\label{app:ablation}
We specified and ran the ablation after the frozen final test, on the public-calibration and development records only; no final-test or paced record is used. Its configuration, including the decision rule, was fixed before the run. Groups, seeds, and paired noise streams are those of the development study, and the rows shared with that study reproduce its archived metrics to within $10^{-13}$.

\paragraph{Motivation.}
Across records, the constant discrete cosine transform (DCT) mode carries a median of about $94$--$95\%$ of a window's log-power spectral density (log-PSD) energy, and modes $1$--$6$ carry about $98$--$99\%$ of the remainder. None of the five random masks contains the constant mode, so the original controls do not separate its contribution from that of the other low-order modes.

\paragraph{Design.}
A \emph{DC-forced} random mask replaces the last index of an original random mask by the constant mode; each of the five has its own public clipping radius. We also evaluate fixed prefixes with $k=1,\ldots,7$, each with its own public radius, and repeat the public search of \cref{eq:public-objective} with $k\in\{1,\ldots,6\}$ added to the grid. A development point qualifies if, compared with the median of the five DC-forced masks, the structured prefix ($k=7$) reduces median relative $L_2$, median band error, and median mean-centered relative $L_2$ by at least $5\%$ each, and wins jointly against at least four of the five masks. Mean-centered relative $L_2$ is \cref{eq:rel-l2} applied after subtracting each curve's mean.

\begin{table}[H]
\centering
\caption{Post-hoc constant-mode ablation on the primary development cohort. Columns three and four are fractional reductions of the structured prefix ($k=7$) relative to the median of five DC-forced random masks, and column five counts its joint wins against them; positive values favor the structured prefix. The last four columns give the dimension selected by public calibration over the extended grid, and the median relative $L_2$ for $k=7$, for the constant mode alone ($k=1$), and at the selected dimension.}
\label{tab:ablation}
\small
\resizebox{\ifdim\width>\textwidth\textwidth\else\width\fi}{!}{%
\begin{tabular}{rrrrcccrrr}
\toprule
 & & \multicolumn{3}{c}{Versus DC-forced masks} & & & \multicolumn{3}{c}{Median relative $L_2$} \\
\cmidrule(lr){3-5}\cmidrule(lr){8-10}
$m$ & $\eps$ & Rel-$L_2$ & Band & Joint wins & Qualifies & Selected $k$ & $k=7$ & $k=1$ & At selected $k$ \\
\midrule
10&0.5&$-$0.025&$-$1.343&0/5&No&1&5.032&1.338&1.338\\
10&1&$-$0.021&$-$1.208&0/5&No&1&2.369&0.733&0.733\\
10&2&$-$0.004&$-$1.190&0/5&No&1&1.218&0.445&0.445\\
10&4&0.038&$-$0.454&0/5&No&1&0.630&0.288&0.288\\
10&8&0.148&0.130&5/5&Yes&2&0.360&0.259&0.176\\
20&0.5&$-$0.021&$-$1.384&0/5&No&1&2.402&0.676&0.676\\
20&1&$-$0.004&$-$1.123&0/5&No&1&1.215&0.378&0.378\\
20&2&0.049&$-$0.473&0/5&No&1&0.629&0.295&0.295\\
20&4&0.193&0.190&5/5&Yes&2&0.316&0.264&0.158\\
20&8&0.395&0.573&5/5&Yes&2&0.178&0.239&0.107\\
40&0.5&$-$0.017&$-$0.720&0/5&No&1&1.317&0.433&0.433\\
40&1&0.018&$-$0.261&0/5&No&1&0.657&0.288&0.288\\
40&2&0.205&0.157&5/5&Yes&2&0.298&0.259&0.169\\
40&4&0.423&0.571&5/5&Yes&2&0.164&0.242&0.101\\
40&8&0.631&0.765&5/5&Yes&3&0.094&0.237&0.073\\
\bottomrule
\end{tabular}}
\end{table}

\paragraph{Result.}
The structured prefix qualifies at six of the fifteen points on the primary development cohort (\cref{tab:ablation}) and at the same six points on the secondary cohort. At the frozen point $(m,\eps)=(40,1)$, it has $1.8\%$ lower relative $L_2$ than the DC-forced median ($0.657$ versus $0.669$) but $26\%$ higher band error ($2.228$ versus $1.767$), with no joint wins; the constant mode alone gives relative $L_2$ $0.288$. The $k=1$ prefix has lower relative $L_2$ than $k=7$ at twelve of the fifteen points; $k=7$ is better only at $(20,8)$, $(40,4)$, and $(40,8)$. With the extended grid, public calibration selects $k=1$ at nine points, $k=2$ at five, and $k=3$ at one, and the resulting release has lower relative $L_2$ than $k=7$ at all fifteen points. Where it selects $k\geq2$, relative $L_2$ is $32$--$69\%$ lower than for the constant mode alone.

The ablation code and configuration were added on 25 November 2026 and run on a Linux cluster with Python~3.12.13. On that machine, rerunning the four original scripts reproduced the archived raw metrics to a maximum absolute difference of $1.1\times10^{-11}$.

\end{document}